\documentclass[preprint]{elsarticle}

\usepackage{bm}% bold math
\usepackage{lineno,hyperref}
\modulolinenumbers[2]

\usepackage{graphicx}% Include figure files
\journal{Computer Physics Communications}

\usepackage{amsmath,amssymb}
\usepackage{float}
\usepackage{xcolor}

\usepackage{changes}
\definecolor{acolour}{RGB}{0, 0, 255}
\definechangesauthor[name={Aleksei}, color={acolour}]{avi}

\def\be{\mathbf{e}}

\begin{document}

\title{Fast and Robust Algorithm for Energy Minimization of Spin Systems Applied in an Analysis of High Temperature Spin Configurations in Terms of Skyrmion Density}

% -----------------------------------
\author[UI,SPBSU]{A.V.~Ivanov}
\author[SPBSU,ITMO]{V.M.~Uzdin}
\author[UI,AU]{H. J\'{o}nsson}

\address[UI]{Science Institute and Faculty of Physical Sciences, University of Iceland VR-III,107 Reykjav\'{\i}k, Iceland}
\address[SPBSU]{Department of Physics, St.\,Petersburg State University, 198504, St.\,Petersburg, Russia}
\address[ITMO]{Faculty of Physics and Engineering, ITMO University, 197101, St. Petersburg, Russia}
\address[AU]{Department of Applied Physics, Aalto University, FIN-00076 Espoo, Finland}

% -----------------------------------

%\date{\today}% It is always \today, today,
             %  but any date may be explicitly specified

\begin{abstract}
An algorithm for the minimization of the energy of magnetic systems is presented and applied to the analysis of thermal configurations of a ferromagnet to identify inherent structures, i.e. the nearest local energy minima, as a function of temperature. Over a rather narrow temperature interval, skyrmions appear and reach a high temperature limit for the skyrmion density. 
In addition, the performance of the algorithm is further demonstrated in a self-consistent field calculation of a 
skyrmion in an itinerant magnet.
The algorithm is based on a geometric approach in which the curvature of the spherical domain is taken into account and as a result the length of the magnetic moments is preserved in 
every iteration. In the limit of infinitesimal rotations, the minimization path coincides with that obtained using damped spin dynamics while the use of limited-memory quasi-newton minimization algorithms, such as the limited-memory Broyden-Fletcher-Goldfarb-Shanno (LBFGS) algorithm, significantly accelerates the convergence. 
\end{abstract}

%\keywords{Suggested keywords}%Use showkeys class option if keyword
                              %display desired
\maketitle

%-------------------------------------------------

\section{\label{sec:intro} Introduction}

Theoretical characterization of magnetic materials typically start with the identification of stable and/or metastable states of the system.
These correspond to local minima on the energy surface, i.e. the energy as a function of the variables corresponding to the various 
degrees of freedom.
At low temperature, a system in thermodynamic equilibrium is characterized by the global minimum on the energy surface.
At a higher temperature, it explores regions of the energy surface favored more by entropy. 
For example, when a ferromagnetic material is heated up, thermal fluctuations in the spin configurations can bring the system into states
with local non-collinear ordering such as skyrmions\cite{Bogdanov1994,Fert2013,Nagaosa2013}.
This has been proposed theoretically based on Heisenberg-type model~\cite{Koshibae2014}
and demonstrated in experiments where short laser pulses briefly heat up the 
system followed by rapid cooling.\cite{Vallobra2018}

High temperature configurations of the spins can be characterized by 
identifying the nearest local minima on the energy surface, i.e. minima that are connected to the high temperature configurations by energy minimization.
The minimization provides a mapping between any configuration of the spins and a local minimum on the energy surface.
The configuration at the local minimum is directly related to the given configuration in the sense that no energy barrier separates the two.
Such local minima on the energy surface are referred to as `inherent structures' of the system at a finite temperature.
Analogous mapping is often used to characterize atomic configurations of liquids\cite{Stillinger86,Jonsson88}.

In the present article, a fast method for minimizing the energy of magnetic systems is presented and applied in a study of the 
inherent structures of a ferromagnet over a wide range in temperature.
A search for a local minimum on the energy surface requires an efficient algorithm for minimizing the energy. 
For magnetic systems, the minimization of the energy is challenging because it involves moving on a curved manifold.
The energy of a system of $N$ spins is effectively a function of the orientation of the spin vectors
\begin{equation}
E = E(\be_1, \be_2 .. \be_N)
\end{equation}
since the length of the spin vectors is either fixed as in a Heisenberg-type Hamiltonian, or it is
treated as a fast variable within an adiabatic approximation~\cite{Antropov1996}
in density functional theory (DFT)~\cite{hohenberg1964,kohn1965} or semi-empirical model Hamiltonian calculations such as the non-collinear Alexander-Anderson (NCAA) model~\cite{Bessarab2014}.
Well-established minimization algorithms that are widely used in atomic simulations to determine the location of atoms at an energy minimum 
cannot be applied to spin systems without significant modifications. 
It is possible to use Cartesian coordinates for the spin vectors, 
$\be_i = (e_{ix}, e_{iy}, e_{iz})$,
but then either a normalization constraint needs to be added,
$|\be_i|^2 = 1$ for $i=1...N$, 
by introducing Lagrange multipliers and thereby increasing the dimensionality of the problem,  
or a renormalization of the spin vectors needs to be introduced~\cite{Cohen1989,Fischbacher2017}.
Alternatively, one can use a spherical coordinate system or a stereographic coordinate system, 
but then a special treatment is required near the poles, 
$\theta = 0$ and $\pi$~\cite{Berkov1996,Bessarab2015,Rybakov2015}.
The choice of coordinate system, such as Cartesian, spherical, or stereographic, can influence both the numerical accuracy 
and the computational effort needed to reach convergence. 
%We leave this comparison for a later study.

An alternative approach is to use an orthogonal optimization algorithm~\cite{Edelman1998}.
This approach has been successfully applied to a variety of problems, 
including signal processing~\cite{Abrudan2009} and electronic structure calculations.~\cite{Hutter1994, VanVoorhis2002, Lehtola2014, Lehtola2016}
The advantage of the orthogonal spin optimisation is that it takes into account the curvature of the manifold on which the energy is defined and, therefore,  
can be combined with algorithms developed for unconstrained minimisation such as conjugate gradient algorithms and quasi-Newton methods.
Here, we apply it to spin systems and show that orthogonal spin optimization (OSO) employed 
in combination with the limited-memory Broyden-Fletcher-Goldfarb-Shanno (LBFGS) method~\cite{Nocedal2006}
including inexact line search provides fast and stable convergence for large spin systems where the inherent structures reveal the presence of magnetic skyrmions.
This OSO-LBFGS algorithm outperforms commonly used 
approaches based on the deterministic Landau-Lifshitz (LL) equation.~\cite{Landau1935}

The article is organised as follows. 
In section~\ref{sec:method}, the orthogonal optimization algorithm is presented. 
In section~\ref{sec:numtest}, results of numerical tests are presented where the orthogonal optimization is carried out in combination with the LBFGS method, the
Fletcher–Reeves nonlinear conjugate gradient (CG) method~\cite{Nocedal2006}  or the
velocity projection optimization (VPO) method~\cite{Jonsson1998,Bessarab2015} algorithms.
Performance is also compared with two LL dynamics approaches, the damped LL and dissipative LL dynamics, as well as with the spin normalization (SN) algorithm of Cohen {\it et al}~\cite{Cohen1989} 
combined with Polak-Ribi\`ere nonlinear conjugate gradient (CG) method.
The SN method is commonly used in micromagnetics simulations~\cite{Fischbacher2017,Exl2019}.
Finally, discussion and conclusions are presented in section~\ref{sec:conclusion}. 
Details of the implementations of the orthogonal optimization algorithm are presented in the Appendices. 
The OSO-LBFGS algorithm has been implemented in the LAMMPS~(https://lammps.sandia.gov) software~\cite{Plimpton1995, Tranchida2018}  
and in the Spirit software~\cite{Mueller2019}.
%well as in the SpinMin python software~\cite{SpinMin}. 
The code as well as initial and final spin configurations of the calculations presented here are available in a GitLab repository.~\cite{SpinMin}

% ------------------------------------------------------------------------

\section{\label{sec:method} Methodology}

\begin{figure}
\includegraphics[width=\columnwidth]{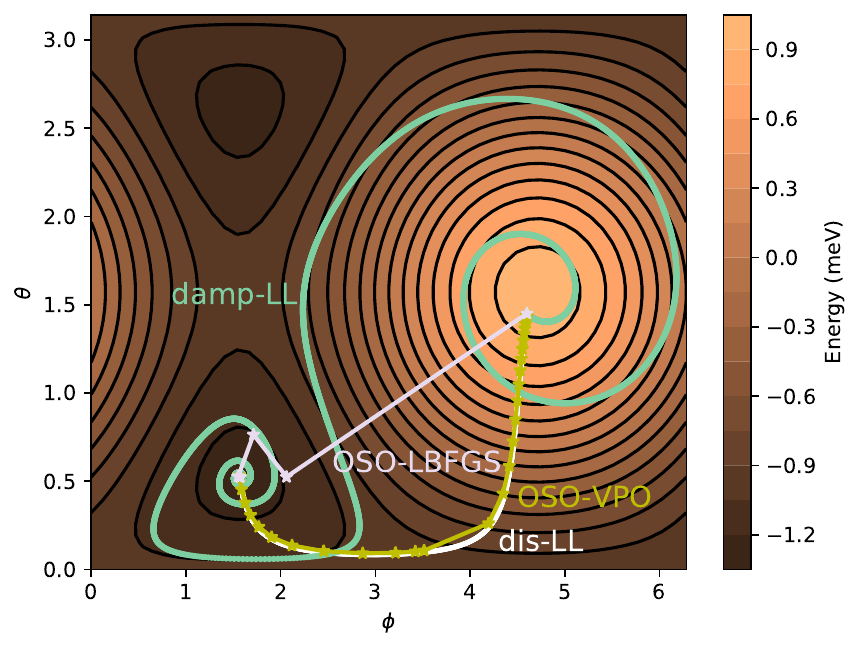}
\caption{
\label{fig:fig1} 
Energy surface of a single spin with easy-axis anisotropy in an external magnetic field: E = $-B \sin(\phi)\sin(\theta) - K \cos^2(\theta)$, $B=K=1$ meV.
The angles $\theta$ and $\phi$ give the orientation of the spin. Several algorithms for finding the energy minimum starting from a point near the maximum are compared. 
The `damp-LL' trajectory is obtained from the damped Landau-Lifshitz equation~\eqref{eq: DLL}.
The `dis-LL'  trajectory is obtained by including only the dissipation term, Eq.~\eqref{eq: LL-dis}, and corresponds to a steepest descent algorithm. The damping parameter is 0.19.
The OSO-LBFGS and OSO-VPO paths correspond to orthogonal spin optimization, Eq.~\eqref{eq: urot}, combined with limited-memory 
Broyden-Fletcher-Goldfarb-Shanno algorithm with inexact line search 
and velocity projection optimization algorithms, respectively.
The OSO-LBFGS clearly outperforms the other approaches as it requires significantly fewer iterations. Stars on the paths correspond to iteration steps of the OSO algorithms.
}
\end{figure}

First, the OSO algorithm is introduced and its implementation as a minimisation method. 
Then, two commonly used minimisation methods based on LL dynamics are discussed for comparison.
The methods are illustrated and compared using calculations for a simple two-dimensional system using the energy surface shown in Fig.~\ref{fig:fig1}.
 
Consider a system consisting of N interacting spins and let $\{\be_i'\}_{i=1}^{N}$ be a reference orientation of the spins. 
Then, {\it any} spin configuration, $\{\be_i\}_{i=1}^{N}$, can be obtained from this reference 
by applying an orthogonal transformation
\begin{equation} \label{eq: urot}
\be_i = U^{i} \be_i'.
\end{equation}
The objective is to find the set of orthogonal matrices that transform the reference spin configuration to the minimum energy configuration. 
Since orthogonal matrices must satisfy orthonormality constraints, ${U^{i}}^{T}U^i = I$, it is convenient to parametrise them using exponentials of
% {\it In order to enforce the 
% normalization
% constraint, 
% the unitary matrices can be parametrised as exponentials} of
skew-symmetric matrices, $A_{i}$:
\begin{equation}
U^{i}  = e^{-A_{i}},
\end{equation}
where
\begin{equation}
A_{i} = 
\begin{pmatrix}
0& a_{12}^{i}& a_{13}^{i}\\
-a_{12}^{i}& 0& a_{23}^{i}\\
-a_{13}^{i}& -a_{23}^{i}& 0\\
\end{pmatrix}.
\end{equation}
Given that the reference spin vectors satisfy the normality constraints $|\be_i'|^2 = 1$, 
the transformed spin vectors $\{\be_i\}_{i=1}^{N}$ also satisfy the normality constraints for any set of $\{ A_i\}_{i=1}^{N}$.  

Skew-symmetric matrices form a linear space and the minimum of the energy can be found as
\begin{equation}
E_{min} = \min_{\vec a \in \mathbb{R}^{3N}} F(\vec a),
\end{equation}
where $ \vec a$ is a $3N-$dimensional vector
\begin{equation}
\vec a = (a^{1} _{12}, a^{1} _{13}, a^{1} _{23}, ...,a^{i} _{12}, a^{i} _{13}, a^{i} _{23}, ..., a^{N} _{12}, a^{N} _{13}, a^{N} _{23}),
\end{equation}
$i$ refers to the spin site, and 
\begin{equation}
F(\vec a) =  E(e^{-A_1}\be_1', e^{-A_2}\be_2'  \dots ,e^{-A_N}\be_N' ) .
\end{equation}
With this formulation, the energy can be minimized as a function of a $3N$-dimensional vector instead of a minimization
with respect to the spin orientations subject to normality constraints. 
In practice, the reference spin orientation is an initial configuration, chosen in the present application from a finite temperature simulation.
It can also represent a guess for the minimum energy spin configuration in order to
obtain convergence with as few iterations as possible,
or it can be chosen at random 
when local minima on the energy surface are being sampled.

The energy gradient with respect to $\vec a$ needs to be evaluated in order to minimize the energy in an efficient way.
Within the OSO approach, the gradient is
 \begin{equation}\label{eq: oso-gr}
 g_{\alpha \beta}^{i} := \frac{\partial F}{\partial a^{i}_{\alpha \beta}} = \left(\int_0^1 e^{t A^{i}} T^{i} e^{-t A^{i}} \,dt \right)_{\alpha\beta},
 \end{equation}
where the matrix $T^{i}$ is 
\begin{equation}\label{eq:T_matrix}
T^{i} =
\begin{pmatrix}
0& t_{iz}& -t_{iy}\\
-t_{iz}& 0& t_{ix}\\
t_{iy}& -t_{ix}& 0
\end{pmatrix}, \quad \vec t_{i} =  \be_{i} \times \frac{\partial E}{\partial \be_{i}} .
\end{equation}
This equation can be obtained from the chain rule of differentiation and the definition of the first directional derivative of the matrix exponential~\cite{Najfeld1995}.

% -----

An iterative minimization can be carried out in two ways, either
\begin{equation}
\label{eq: small_rotation}
\be_i^{(k+1)} = e^{-A\,^{(k)}_i} \cdots e^{-A\,^{(1)}_i} e^{-A\,^{(0)}_i} \be'_i, \forall i \in {1, 2, .., N}
\end{equation}
or
\begin{equation}
\label{eq: large_rotation}
\be_i^{(k+1)} = e^{-(A\,^{(k)}_i + \dots + A\,^{(1)}_i + A\,^{(0)}_i)} \be'_i, \forall i \in {1, 2, .., N} .
\end{equation}
The advantage of Eq.~\eqref{eq: small_rotation} is that 
each rotation is small and 
the gradient vector at each iteration can be calculated as
\begin{equation}\label{eq: gradient_small_rotation}
    g_{\alpha\beta}^{i} = T^{i}_{\alpha\beta}
\end{equation}
since the reference spins are updated at each iteration in Eq.~\eqref{eq: small_rotation}
\begin{equation}
\label{eq: small_rotation2}
\be_i^{(k+1)} = e^{-A\,^{(k)}_i} \be_i^{(k)}, \forall i \in {1, 2, .., N} .
\end{equation} 
Both iteration processes have been implemented and tested. We have not found a significant difference in the number of energy and gradient evaluations while the evaluation of gradients in Eq.~\eqref{eq: small_rotation} requires less computational time. In order to reduce further the computational effort, 
the Cayley transformation could be used instead of the matrix exponential as it does not require calculation of trigonometric functions~\cite{Lewis2003}.

The skew-symmetric matrices used in the matrix exponential have a physical interpretation for rotations in spin systems. Namely, the quantity 
\begin{equation}
\theta_i = \sqrt{-\frac{1}{2}\mathrm{Tr}\left[{\left(A^i\right)}^2\right]}=\sqrt{\left(a_{12}^{i}\right)^2 + \left(a_{13}^{i}\right)^2 + \left(a_{23}^{i}\right)^2}
\end{equation}
defines the rotation angle of the $i$\textsuperscript{th} spin around the axis
\begin{equation}
 \mathbf{r}_i = \frac{1}{\theta_i} (-a^{i}_{23}, a^{i}_{13}, -a^{i}_{12})^{T} .
\end{equation}
Using this interpretation, one can calculate the matrix exponential using Rodrigues' formula~\cite{Murray1994} as
\begin{equation}\label{eq:rod}
e^{{A}^i}= e^{\theta {A'}^i} = I + \sin(\theta){A'}^i + \left(1-\cos(\theta)\right) {\left({A'}^i\right)}^2,
\end{equation}
where ${A'}^i= A^i/\theta$ (see Appendix~\ref{app:um} for details).  
This opens the possibility of introducing a cutoff step length along the search direction in the minimization algorithm: 
if the root-mean-square angle $\theta_{rms} = \sqrt{\sum_{i=1}^{N}\theta_i^{2}/N}$ is larger than some threshold angle, $\theta_{max}$, 
then the rotation angles are rescaled but the rotation axes kept the same by carrying out the transformation:
\begin{equation}
\theta_i \leftarrow \theta_i \frac{\theta_{max}}{\theta_{rms}} \quad \text{if } \frac{\theta_{max}}{\theta_{rms}} < 1, 
\end{equation}
This makes the algorithm stable and can be particularly important in the beginning of a minimization if the line search procedure is not performed. In practical implementation, instead of calculating the matrix exponential~\eqref{eq:rod} one can calculate the action of the matrix exponential on a vector as
\begin{equation}\label{eq:rotee}
 e^{-A^i} \be_i = 
 \cos\theta_i \be_i + 
 \sin\theta_i \mathbf{r}_i \times \be_i + 
 (1-\cos\theta) \mathbf{r}_i \cdot \be_i \, \mathbf{r}_i.
\end{equation}
% --------------------------------------------------

By defining the energy of the spin system as a function of variables in linear space and having the corresponding expression for the energy gradient,
well-established algorithms can be used to carry out the energy minimization. 
%We will refer to this approach as orthogonal spin optimization (OSO) and 
We have tested three different minimization methods for finding the coefficients in the vector $\vec a$, namely the LBFGS, CG and VPO methods.
The details of our implementation of the minimization algorithms and the calculation of the orthogonal matrix and gradient of the energy are given in~\ref{app:OSO-SmallRot}-\ref{app:um}.

% --------

In simulations based on electronic structure calculations such as the DFT or NCAA methods as well as in calculations of large systems including many spins and long range interactions in a Heisnberg-type Hamiltonian, the evaluation of the energy and its gradient, i.e. the effective field, 
becomes the most time-consuming operation. 
Therefore, the computational effort in a minimization calculation is quantified here by the number of evaluations of the energy and its gradient.

% -----------

%  2d test problem

A simple test problem involving two degrees of freedom is characterized by the energy surface shown in Fig.~\ref{fig:fig1}.
The results of two types of OSO minimisation approaches are shown, the OSO-LBFGS and OSO-VPO algorithms. 
%
% ---------------------------------- figure 1 ------------------------------
% -----------------------------------------------------------------------------
%
The OSO-VPO calculation follows essentially a steepest descent path, but because of the acceleration that occurs when the gradient
points in a similar direction in consecutive iterations, the steps size along the path varies.~\cite{Jonsson1998,Bessarab2015} 
The method is quite stable but suffers from
a slowing down as the magnitude of the gradient becomes small and the path curves near the minimum, 
leading to a small step size, see Fig.~\ref{fig:fig1}.

The combination of OSO with the LBFGS algorithm is more efficient. As can be seen in Fig.~\ref{fig:fig1}, the path taken from the initial
point to the minimum is rather direct and involves much fewer iterations than for the OSO-VPO method.

% -------------
%  LL dynamics

In many cases the minimisation of the energy of a magnetic system is carried out using damped spin dynamics obtained by
integrating the deterministic Landau-Lifshitz (LL) equation~\cite{Landau1935}
\begin{equation}
\label{eq: DLL}
\frac{d \be_i}{d t} = - \frac{\gamma}{M_i}\be_i \times \vec{b}_{i}^{ \rm eff} - \alpha\frac{\gamma}{M_i} \, \be_i \times \be_i \times \vec{b}_{i}^{ \rm eff},
\end{equation}
where $\be_i$ is a unit vector defining the direction of the magnetic moment of the $i$\textsuperscript{th} spin, 
$\gamma$ is the gyromagnetic ratio, $M_i$ is the length of the magnetic moment, $\vec{b}_{i}^{ \rm eff}$ is the effective field
\[\vec{b}_{i}^{ \rm eff} = -\frac{\partial E }{\partial \be_i}\]
and $\alpha$ is the damping parameter representing some coupling of the system to a heat bath.
Along the trajectory obtained from this equation, the energy of the system leaks out through the damping term, eventually reaching a local minimum.
We will refer to this as the `damp-LL' minimisation method.
As can be seen from  Fig.~\ref{fig:fig1}, the calculation gives a long and winding path from the initial point to the minimum. 
This is clearly not an efficient minimisation method.  
Nevertheless, it is often used in computational studies of magnetic systems.
Furthermore, since the energy leaks only gradually out of the system,
it is possible that an energy barrier, lower than the energy of the initial point, is traversed during the minimisation, so the method
may not provide a mapping to the inherent structure. 

A more direct path to the minimum can be obtained by eliminating the oscillatory term in Eq.~\eqref{eq: DLL} 
and including only the dissipative term
\begin{equation}\label{eq: LL-dis}
\frac{d \be_i}{d t} = - \alpha \frac{\gamma}{M_i} \, \be_i \times \be_i \times \vec{b}_{i}^{ \rm eff} .
\end{equation}
We will refer to this as the `dis-LL' method and a calculated path for the two-dimensional test problem is shown in Fig.~\ref{fig:fig1}.
The path obtained corresponds essentially to steepest descent on the energy surface, similar to OSO-VPO,
but requires more iterations.  
Near the minimum especially, where the magnitude of the gradient is small, the method requires many evaluations for the final approach,
thereby reducing its efficiency.

% -------

The orthogonal optimization method can be shown to reduce to the steepest descent and dis-LL in the limit of small rotations.
For a small rotation, $\Vert A^{i} \Vert \ll 1$ for all $i = 1 \dots N$, the gradient becomes
$g_{\alpha \beta}^{i} = T^{i}_{\alpha\beta}$ 
and a rotation in the steepest descent direction at $A^{i} = 0$ is
\begin{equation}
\be_{i}= e ^{\lambda T^{i}} \be_i' \approx (I + \lambda T^{i}) \be_{i}' = 
\be_{i}' + \lambda \be_{i}'  \times \be_{i}'  \times \frac{\partial E}{\partial \be_{i}'} .
\end{equation}
If one chooses $\lambda = \gamma \alpha \Delta t /M_i$ and $\Delta t \rightarrow 0$, then the equation above transforms to
\begin{equation*}
\frac{d \be_i' }{d t} = \alpha \frac{\gamma}{M_i} \be_i'  \times \be_{i}'  \times \frac{\partial E}{\partial \be_{i}'}
\end{equation*}
which is the same as Eq.~\eqref{eq: LL-dis}.
Thus, in the limit of small rotations, where the orthogonal optimisation becomes equivalent to a steepest descent algorithm,
it corresponds to the dis-LL trajectory.
%~\eqref{eq: DLL}. 
%  ----------------------------------------------------------------------------------------------------------------------------------------------------------------

\section{\label{sec:numtest} Performances tests}
The performance tests are carried out using an extended Heisenberg-type Hamiltonian with short-range interaction as well as self-consistent mean field (SCF) calculations within NCAA model.  
The computational complexity of an extended Heisenberg-type Hamiltonian scales linearly with number of spins in the system, 
while the SCF calculations scale as the third power of the number of spins in conventional implementations.

\subsection{Skyrmion lattice in an extended Heisenberg-type Hamiltonian.}
We will use Heisenberg-like Hamiltonian with short-range interaction which describes magnetic moments arranged on a planar square  lattice. 
\begin{equation}\label{eq:heisenberg}
E =-\frac{1}{2}\sum_{<j,k>}(J_{jk} {{\bf e}_j\cdot {\bf e}_k+\vec{D}_{jk}\cdot({\bf e}_j \times {\bf e}_k))}
-K \sum_j {e^2_{j,z}}- M \vec{H} \sum_j {\bf e}_j,
\end{equation}
 where $ \bf e_j $ is a three-dimensional vector of unit length along the magnetic moment on site {\it j}, $\it M$ is the  magnitude of the magnetic moments, which is presupposed to be the same for all sites, $\it J_{jk}$ is the exchange parameter, which we assume to be non-zero ($\it J_{jk}=J$) only for the nearest-neighbour sites, $\vec{D}_{jk}$ is the Dzyaloshinsky-Moriya vector directed  along the line connecting atomic sites {\it i} and {\it j} and {\it K} is the anisotropy parameter. Anisotropy vector points perpendicular to the lattice plane. The summation $<i, j>$ runs over all pairs of nearest neighbour sites.
 
The supercell consists of either 20$\times$20 or 40$\times$40 spins subject to periodic boundary conditions. 
The parameters in the Hamiltonian are in this case 
$J = 10$ meV, $D = J / 2$ (where D is the length of the DM vector), $K=0$,
and  the magnetic field is perpendicular to the monolayer with $M |\vec{H}| = J / 5$. 
The ground state of the system corresponds to ferromagnetic ordering with energy of -22 meV per spin. 
The energy surface has multiple local minima corresponding to skyrmions of varying density.
The skyrmionic states found in the 20x20 and 40x40 systems have energy -21.94 and -21.98 meV per spin, respectively.

% ---------------------------------- figure 2 ------------------------------
\begin{figure}
\includegraphics[width=0.9\columnwidth]{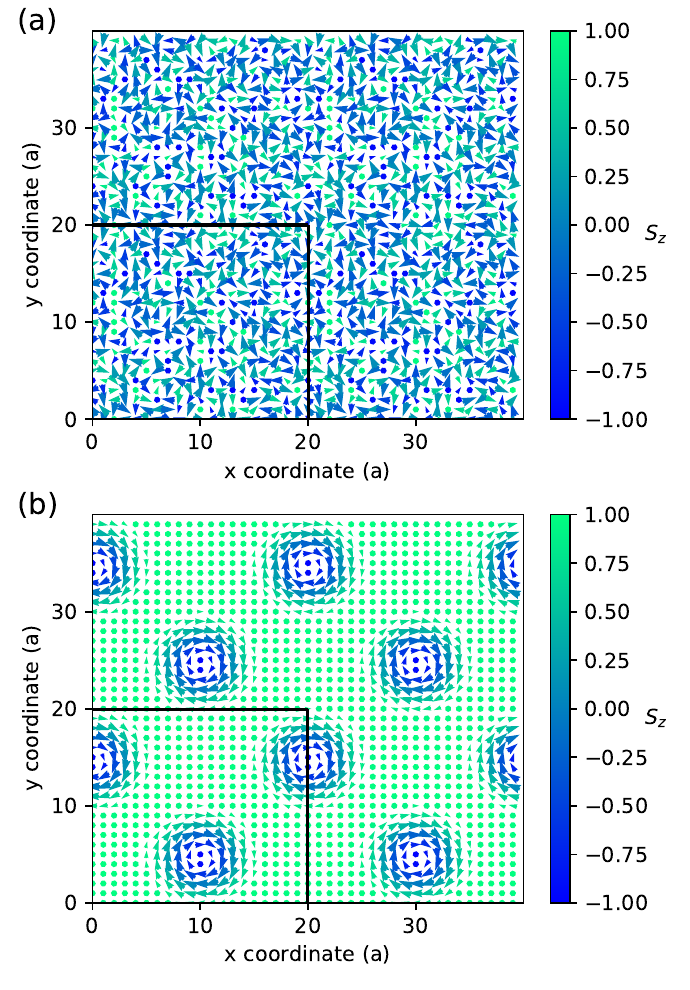}
\caption{\label{fig:fig2}
A 20$\times$20 lattice of spins subject to periodic boundary conditions 
(the simulation cell is marked by solid black lines, but three periodic images of the system are also shown). 
Coordinates are given as a multiple of the lattice constant.
(a): Initial state where the spins are randomly oriented.
(b): A local minimum on the energy surface corresponding to two magnetic skyrmions in the simulation cell,
found using the OSO-LBFGS minimisation algorithm. 
The three other algorithms tested gave equivalent skyrmion states with the same energy but in some cases different
location of the skyrmions.
}
\end{figure}
%  ------------------------------------------------------------------------------------------------------

The performance of four algorithms is compared.  The OSO is combined with either VPO, CG or LBFGS algorithms using the gradient expression given in section~\ref{sec:method}. The CG and LBFGS algorithms include inexact line search.~\cite{Nocedal2006}
The fourth algorithm is dis-LL, Eq.~\eqref{eq: LL-dis}, where the SIB algorithm 
is used to generate the trajectory since it preserves the length of the spins even for large time steps.~\cite{Mentink2010}
The damping parameter is taken to be $\alpha$=0.1 and the time step set to 0.7 ps. For a larger time step of 0.8 ps the SIB algorithm diverges.
The initial spin orientations are chosen from a random distribution on a unit sphere. 
Convergence is considered to be achieved when 
the maximum magnitude of the torque acting on the magnetic moments is less than $10^{-5}$ meV, 
that is $\max_i | \vec{t}_i|  < 10^{-5}$ meV. 

% ---------------------------------- figure 3 ----------------------------------
\begin{figure}
\includegraphics[width=0.9\columnwidth]{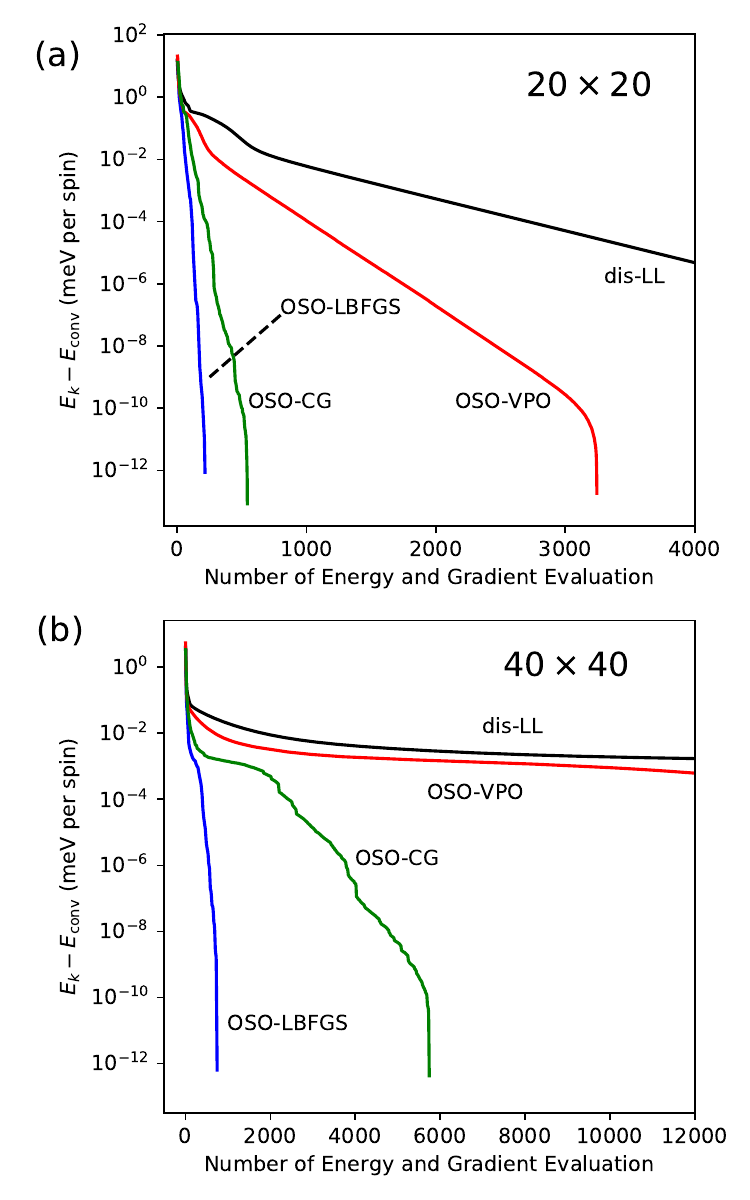}
\caption{\label{fig:fig3} 
 Energy as a function of the number of energy/gradient evaluations using 
 the three different implementations of the orthogonal spin optimization algorithm, LBFGS, CG or VPP methods,
 and the dis-LL dynamics equation integrated using the SIB 
 algorithm.
Convergence is taken as $\max_i | \hat{s}_i \times {\partial E}/{\partial \hat{s}_i} | < 10^{-5}$ meV.
(a): 
Results for the 20$\times$20 lattice depicted in Fig.\ref{fig:fig2}. 
(b):
Results for the 40$\times$40 lattice depicted in Fig.\ref{fig:fig4}. 
}
\end{figure}
% ------------------------------------------------------------------------------------------------

The rate of convergence of the energy is shown in Fig.\ref{fig:fig3}. The OSO-LBFGS algorithm shows the best performance, requiring {\it ca.} 220 
energy/gradient evaluations to converge on the skyrmion state.
The OSO-CG requires about 50\% more evaluations to reach the same level of convergence.
The OSO-VPO is much slower requiring more than 3000 evaluations. 
The dis-LL algorithm has the worst performance requiring  ca. 4000 iterations. 
As one iteration in the SIB algorithm requires two evaluations of the gradient for calculating the predictor and corrector, 
the total number of energy/gradient evaluations becomes {\it ca.} 8000 in the dis-LL calculation.
This shows that large savings in computer time can be achieved by using the OSO-LBFGS in energy minimizations for magnetic systems.
 
%The number of energy and gradient evaluations is not the same as the number of iterations.  In the SIB dynamics algorithm, 
%two evaluations of the effective fields (first in order to calculate the predictor and then in order to calculate the corrector) are required for each iteration. 
While the LBFGS and CG algorithms in principle require only one evaluation of the gradient at each iteration, 
the energy must also be estimated in order to test the strong Wolfe conditions and if these conditions are violated a line search procedure needs to be applied 
(see Appendix~\ref{app:OSO-SmallRot} ). For LBFGS it is in general only necessary to perform the line search in the beginning phase of the minimization.
After a few minimization steps have been carried out, the unity step length along the LBFGS search direction is guaranteed to satisfy the strong Wolfe conditions if 
%all 
earlier steps have satisfied the conditions.~\cite{Nocedal2006} Therefore, most iterations in the LBFGS algorithm require only one evaluation of the energy and the gradient. 
  
% ---------------------------------- figure 4 ------------------------------
\begin{figure}
\includegraphics[width=0.9\columnwidth]{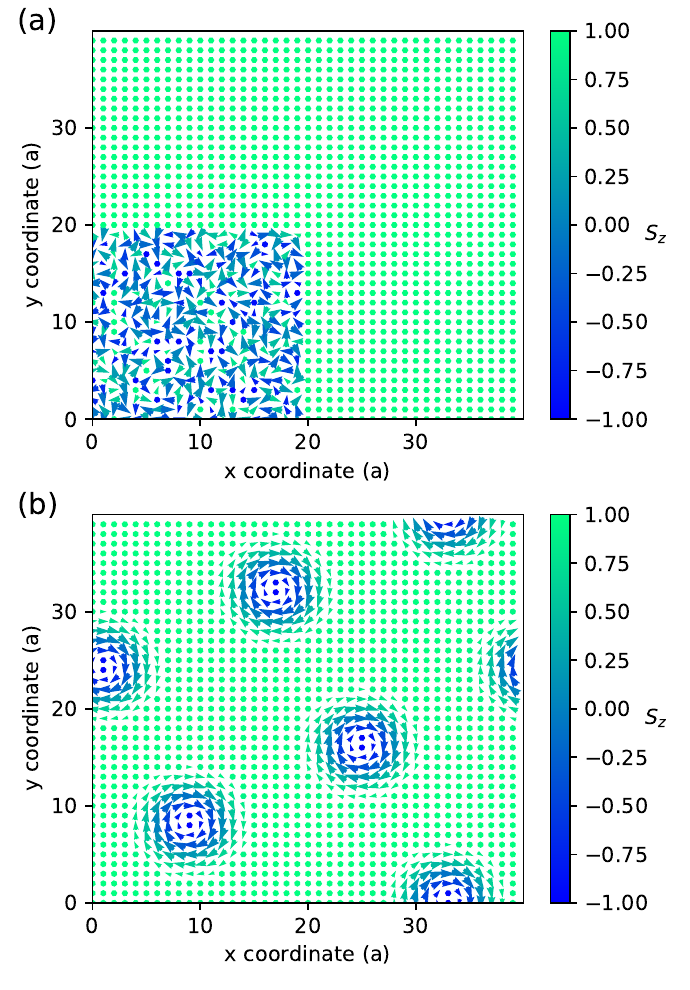}
\caption{\label{fig:fig4}
A 40$\times$40 lattice of spins subject to periodic boundary conditions. Coordinates are given as a multiple of the lattice constant.
(a):  Initial state.
(b): Local minimum obtained with OSO-LBFGS. The three other algorithms tested converge to an equivalent state with the same energy corresponding to five magnetic skyrmions in the
simulation cell.
}
\end{figure}
%  ------------------------------------------------------------------------------------------------------

When the number of degrees of freedom is increased by adding more spins to the simulated system, the number of energy/gradient evaluations needed to 
reach convergence also increases.  This is illustrated by simulating a 40$\times$40 lattice where spins in a 20$\times$20 subsection of the lattice initially have 
random spin orientations, the same as in the calculation of the smaller system shown in Fig.\ref{fig:fig2}(a), while the rest of the spins are oriented as in the ferromagnetic phase, see Fig.\ref{fig:fig4}(a). 
The OSO-LBFGS calculation now requires 750 evaluations to converge to a tolerance of $10^{-5}$ meV, 
while OSO-CG and OSO-VPO require 5700 and 45000, respectively.
 The dis-LL minimization requires {\it ca.} 60000 iterations (120 000 energy/gradient evaluations). 
 Interestingly, all four methods converge on the same local minimum energy spin configuration
 where the density of skyrmions is {\it ca.} half as large as in the smaller 20$\times$20 system.
 This calculation illustrates how the addition of more degrees of freedom, without introducing additional complexity 
 (in that the number of randomly oriented spins is the same as in the calculation illustrated in Fig.\ref{fig:fig2}) 
 increases the number of evaluations required to reach convergence. 
 The damp-LL algorithm convergences to a different minimum and as a result is not included in the comparison.
 Furthermore, the number of iterations exceeds 70000 for convergence.
 
 %-------------------- Table 1 ----------------------
 \begin{table}
 \begin{center}
 \caption{
 \label{tab: tab1} Number of iterations required to reach various convergence criteria,
 defined as tolerance in the magnitude of the gradient of the energy 
 %  HJ change
 % per
 on a 
 spin. One iteration in dis-LL requires two evaluations of the gradient when the SIB algorithm is used. 
 }
% \begin{ruledtabular}
 \begin{tabular}{c c c c c}
 %Convergence 
 Tolerance (meV)& dis-LL & VPO & CG & LBFGS \\
 \hline
%$10^{-4}$ & 45287 & 34313 & 4558 & 716\\
$10^{-4}$ & 45300 & 34300 & 4560 & 715\\
 %$5\cdot10^{-5}$ & 49638 & 37586 & 4848 & 739\\ 
 $5\cdot10^{-5}$ & 49600 & 37600 & 4850 & 740\\ 
 %$10^{-5}$ & 59754 & 45169 & 5746 & 747
 $10^{-5}$ & 59750 & 45200 & 5750 & 750
 \end{tabular}
% \end{ruledtabular}
\end{center}
 \end{table}
 %-----------------------------------------------------
 Table~\ref{tab: tab1} shows how the number of energy/gradient evaluations depends on the convergence criterion. 
 The LBFGS requires only slightly more iterations when the convergence criterion is decreased by an order of magnitude, while the other algorithms require {\it ca.} 20-30 
 $\%$ more iterations. 
 %-----------------------------------------------------
 
In order to reduce the dependence on the initial random configuration, 40 different random configurations were generated~\cite{random} for which all the tested algorithms converged to the same energy minimum. The average number of energy and gradient evaluations, number of iterations as well as 
elapsed
time are presented in Table~\ref{tab: tab2}. The table also includes results obtained with the SN algorithm implemented with Polak-Rebi\`ere (PR) nonlinear conjugate gradient algorithm. The SN combined with PR algorithm is slower than OSO-LBFSG by a factor of 3-4 but performs better than the OSO-FR algorithm. Note that the ratio of energy/gradient evaluations and number of iteration in the LBFGS algorithm is close to one as the step size of 1 is the natural step-size for quasi-Newton algorithms~\cite{Nocedal2006} while nonlinear conjugate gradient algorithms perform a line search more often.

 %-------------------- Table2 ----------------------
 \begin{table}[H]
 \begin{center}
 \caption{
 \label{tab: tab2} 
 Performance results averaged over 40 initial random configurations in terms of the number of energy and gradient evaluations (no.\,e/g), number of iterations (no.\,iter), and elapsed time in seconds required for various algorithms to converge on the energy minimum. The convergence is deemed to be achieved when the maximum torque acting on any one of the magnetic moments is less then $10^{-5}$ meV. All algorithms converge to the same energy minimum for these 40 initial configurations. The algorithms were implemented using Python and the Numpy package and the calculations performed on a single Intel Xeon Gold 6130 CPU (2.10GHz).}
 \begin{tabular}{c c c c}
 Method & no. e/g & no. iters & 
elapsed time\\
 \hline
SIB-dis-LL & 124309 & 62155 & 147.8\\
OSO-VPO & 47041 & 47041 & 61.0\\
OSO-FR-CG& 4827 & 4333 & 6.2\\
SN-PR-CG & 2542 & 1826 & 4.0\\ 
OSO-LBFGS & 724 & 687 & 1.0 \\
 \end{tabular}
\end{center}
 \end{table}
 %-----------------------------------------------------

% ----------------------------------------------------------------------------------------------------------------------------------

\subsection{Skyrmions in NCAA model}
In the NCAA model, the electronic structure of a 3{\it d} transition metal is approximated by two bands: One representing quasi-localized {\it d}-electrons and the other representing itinerant {\it s}({\it p})-electrons.
In order to describe non-collinear magnetic states, we use a mean field approximation at each site $i$ where a local quantization axis, $z_i$, is chosen to be along the local magnetic moment associated with atom $i$ \cite{Bessarab2014}.
The Hamiltonian  for the d-subsystem in mean field approximation can be written as 
\begin{equation}
\label{eq:AA_MF}
\hat H=\sum\limits_{i,\alpha}E_{i}^\alpha \hat n_{i\alpha}+\sum\limits_{i, j,\alpha,\beta}V_{ij}^{\alpha\beta} \hat d_{i\alpha}^{\dag}\hat d_{j\beta}- \frac{1}{4}\sum\limits_i U_i\left(N^2_i-M^2_i\right),
\end{equation}
where 
\begin{equation}
\label{eq:d_level}
E_i^{\alpha} = E_i^0 + \frac{U_i}{2}\left(N_i-\alpha \cos \theta_i M_i\right),
\end{equation}
\begin{equation}
\label{eq:d_hopp}
V_{ij}^{\alpha\beta} = \frac{U_i}{2}\left(\delta^{\alpha\beta}-1\right)\delta_{ij}\exp \left(-\alpha \mathrm i \phi_i\right)\sin \theta_i M_i + \left(1-\delta_{ij}\right)\delta^{\alpha\beta}V_{ij}.
\end{equation}
Here, $E_i^0$ is a renormalized  energy
of unperturbed {\it d}-levels, $U_i$ is on site Coulomb repulsing and $V_{ij}$ are the hopping parameters that contain both a contribution due to direct exchange between {\it d}-states localized on sites $i$ and $j$ and a contribution from indirect {\it d}-{\it d} coupling through the conduction band. 
The {\it d}-electrons are included explicitly here, while the influence of the itinerant {\it s}({\it p})-electrons is indirectly 
taken into account via the renormalization of model parameters.
The hybridization of {\it s}({\it p}) and {\it d} bands leads to broadening of the {\it d}-band
and the width, $\Gamma$, is assumed to be a parameter in the model.
The polar angle $\theta_i$ and the azimuthal angle $\phi_i$ define the direction of the $i$th magnetic moment
with respect to the laboratory quantization axis, $z$. 
In Eqns.~(\ref{eq:AA_MF})-(\ref{eq:d_hopp}), the number of {\it d}-electrons, $N_i$, and the magnitude of magnetic moment, $M_i$, associated with
one of the   five degenerate {\it d}-orbitals at atom $i$ are
\begin{equation}
\label{eq:N}
N_i = \langle \hat{\tilde{d}}_{i+}^{\dag}\hat{\tilde{d}}_{i+}\rangle + \langle \hat{\tilde{d}}_{i-}^{\dag}\hat{\tilde{d}}_{i-}\rangle=\langle \hat d_{i+}^{\dag}\hat d_{i+}\rangle + \langle \hat d_{i-}^{\dag}\hat d_{i-}\rangle,
\end{equation}
\begin{equation}
\label{eq:M}
\begin{split}
M_i &= \langle \hat{\tilde{d}}_{i+}^{\dag}\hat{\tilde{d}}_{i+}\rangle - \langle \hat{\tilde{d}}_{i-}^{\dag}\hat{\tilde{d}}_{i-}\rangle=\cos \theta_i\left(\langle \hat{d}_{i+}^{\dag}\hat{d}_{i+}\rangle - \langle \hat{d}_{i-}^{\dag}\hat{d}_{i-}\rangle\right)\\
& + \sin \theta_i \left(e^{-\mathrm i \phi_i}\langle \hat{d}_{i+}^{\dag}\hat{d}_{i-}\rangle + e^{\mathrm i \phi_i}\langle \hat{d}_{i-}^{\dag}\hat{d}_{i+}\rangle\right).
\end{split}
\end{equation}
For a given orientation of the magnetic moments, specified 
by the angles $\theta_i$ and $\phi_i$, the magnetic structure is described by a set of self-consistent values of $N_i$ and $M_i$. A detailed description of the self-consistent procedure can be found in \cite{Ivanov2020ncaa}.

\begin{figure}
\begin{center}
\includegraphics[width=0.8\columnwidth]{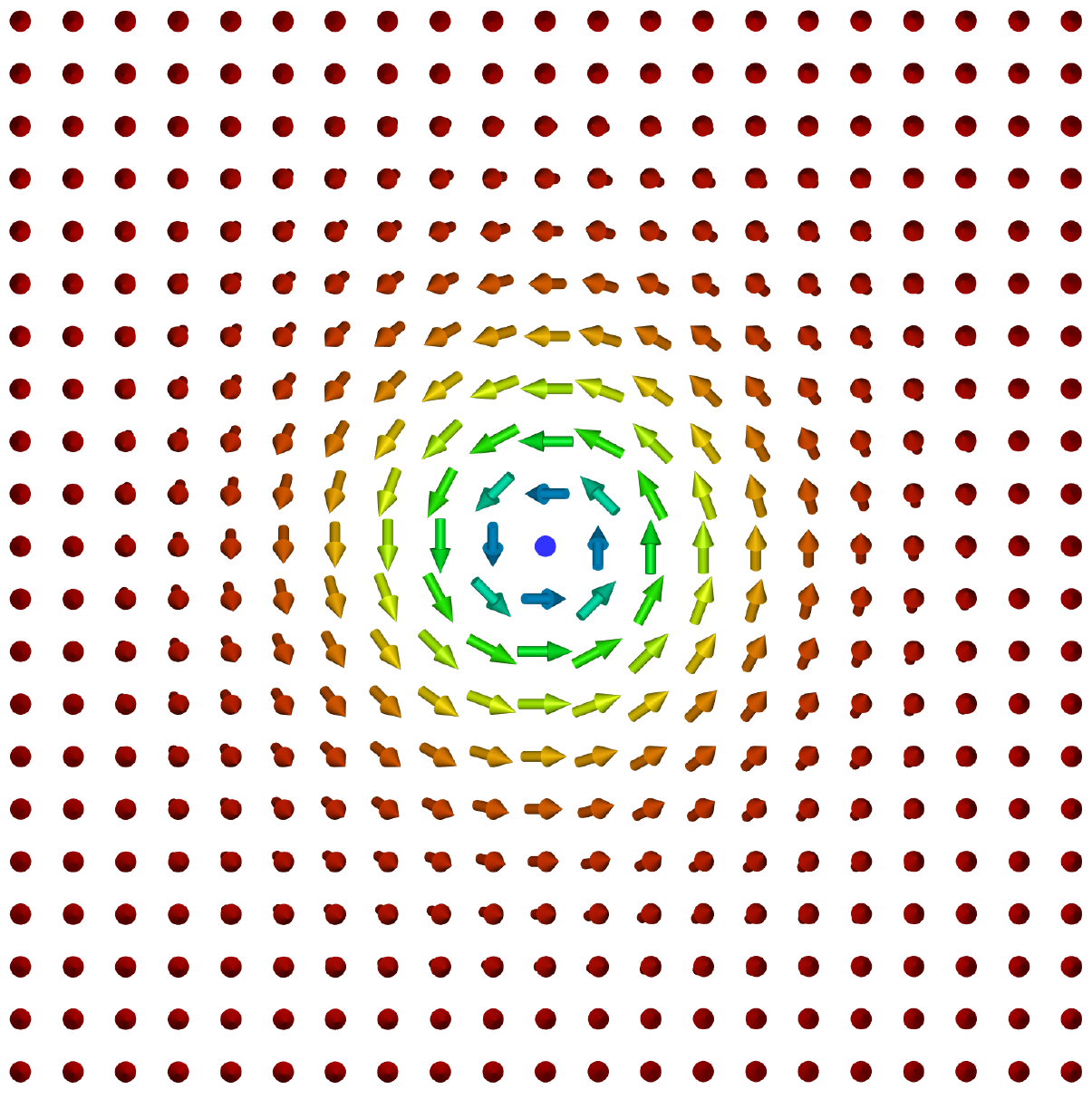}
\caption
{
\label{fig:fig6}
A skyrmion obtained within the NCAA model. Color corresponds to $z$ component of arrows, which represent directions of magnetic moments, ranging from red (+1) to green (0) to blue (-1). The image was generated with the Spirit software~\cite{Mueller2019}
}
\end{center}
\end{figure}

In benchmark calculations the simulation cell consists of 21x21 spins placed on a square lattice subject to periodic boundary conditions. The initial guess for spin orientations is:
\begin{align}
& {\bf e}_i =  \left( \sin(\theta_i) \cos(\phi_i), \sin(\theta_i) \sin(\phi_i),\cos(\theta_i)\right)^T\\
&  \phi_i = {\rm arctan}\left(\frac{y_i-y_0}{x_i-x_0}\right) + \frac{\pi}{2}\\
& \rho_i = \sqrt{(x_i-x_0)^{2} + (y_i-y_0)^2}\\
& \theta_i = \pi \left(1 - \frac{\rho_i}{A} \right), \quad \rho_i < A\\
& \theta_i = 0, \quad \rho_i \geq A
\end{align}
with $A= 5a$ where $a$ is the lattice constant. The magnetic moments were relaxed until the maximum torque dropped than $10^{-6} \Gamma$. At each minimization iteration $M_i$ and $N_i$ were converged with accuracy of $10^{-10}$. The final spin configuration is shown in Fig~\ref{fig:fig6}. 

A comparison of the performance of the various minimization algorithms for this system is presented in Table~\ref{tab: tab3}. The most computationally demanding part is the SCF step which requires calculation of eigenvalues and eigenvectors of the NCAA Hamiltonian. The OSO-LBFGS algorithm converges faster than the other algorithms and requires around 1600 SCF steps which is ca. a third compared with the other algorithms. While SN-PR-CG carries out fewer outer steps then OSO-FR-CG, it is still slower then OSO-FR-CG because it performs more line searches. The total wall-time is proportional to the number of SCF steps for all algorithms which shows that it is indeed the most computational demanding operation and the performance of the algorithm can be quantified by counting the number of SCF steps. 

 %-------------------- Table2 ----------------------
 \begin{table}[H]
 \begin{center}
 \caption{
 \label{tab: tab3} 
Performance of various minimization algorithms in a calculation of a skyrmion in the NCAA model. 
 The number of self-consistent field steps (no.\,SCF steps), number of energy and gradient evaluations (no.\,e/g), number of iterations (no.\,iters), and  elapsed time in seconds
 are given. 
 The algorithms were implemented using Python and the Numpy package and the calculations performed on 8 Intel Xeon Gold 6130 CPUs (2.10GHz).
 }
 \begin{tabular}{c c c c c}
 Method &no.\,SCF steps &no.\,e/g &no.\,iters &  elapsed time \\
 \hline
SN-PR-CG& 4566 & 423 & 285 & 816\\ 
OSO-FR-CG& 4379 & 383 & 328 & 780\\
OSO-LBFGS& 1589 & 126 & 120 & 281 \\
 \end{tabular}
% \end{ruledtabular}
\end{center}
 \end{table}

% ---------------------------------------------------------------------  Fig. 5  -------------------------------------------------

% ----------------------------------------------------------------------------------------------------------------------------------

\section{\label{sec:application} Application: Inherent structures at high temperature}

\begin{figure}
\includegraphics[width=\columnwidth]{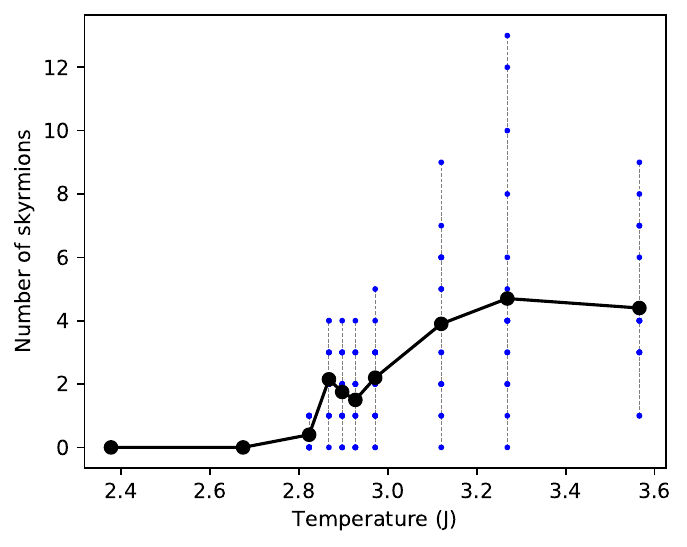}
\caption
{
\label{fig:fig5}
Number of skyrmions found after minimization of the energy starting from configurations where a circular region within the layer has been heated to
a given temperature (given in units of the exchange constant, $J$). The heated region contains 1.7$\times$10$^5$ spins while the simulation box consists of 
2.5$\times$10$^5$ spins subject to periodic boundary conditions. The black curve shows the average number of skyrmions found from 20 uncorrelated initial configurations. 
The dots show the number of skyrmions found for each configuration, thereby indicating the spread. 
}
\end{figure}

We now apply the OSO-LBFGS algorithm to study inherent structures of a ferromagnetic layer as a function of temperature.
As mentioned in the introduction, both theoretical calculations based on a Heisenberg-type model~\cite{Koshibae2014} and experimental studies 
of thin films~\cite{Vallobra2018} have shown that high temperature can
break the ferromagnetic order and introduce skyrmions which remain in the system if it is cooled quickly enough.
The experiments were carried out for a system where skyrmions are large (micron size) and stable at room temperature. 
The simulation of such a large system on the atomic scale is a challenging task. 
Here, we have instead investigated the skyrmion creation due to local heating in a well known system that has been studied extensively, namely the Co/Pt(111) system. 
The values of the parameters of the Hamiltonian  
are chosen to describe N\'eel skyrmions: $J = 29$ meV,  $|\vec{D}_{ij}|=1.5$ meV, $K=0.4$ meV as has been done previously by Rohart and
coworkers~\cite{Rohart2016}. 
The external field is set to zero, $H=0$.
The effect of the dipole-dipole interaction is accounted for effectively to a good approximation by reducing the value of the anisotropy constant to $K=0.293$ meV~\cite{Rohart2016,Lobanov2016}.
The simulated system consists of 250000 spins in the simulation cell arranged on a triangular lattice subject to periodic boundary conditions.
The initial configurations for the energy minimisations are prepared by heating spins in a region within a 
circle of radius 251$a$ (where $a$ is the lattice constant), including 167786 spins, using stochastic Landau-Lifshitz-Gilbert (LLG) dynamics over a period of 1.0 ns in order to reach thermal equilibrium.
A total of 20 statistically uncorrelated configurations are prepared by recording the orientation of the spins at 0.2 ns intervals after the equilibration period. 
The other spins in the system remain frozen in the ferromagnetic arrangement during this preparation phase.
This is repeated for a wide range in temperature. 
For each spin configuration generated in this way, an OSO-LBFGS minimization~(see Appendix~\ref{app:lbfgs}) is carried out with all spin orientations being
free degrees of freedom.
The average number of skyrmions found 
at the minimum energy configuration is reported in Fig.~\ref{fig:fig5} as a function of the temperature from which the initial configuration was obtained. 
The temperature is given in units of the exchange parameter, $J$. 
For temperature below 2.8 $J$ (corresponding to about 800 K for the Co/Pt(111) system) no skyrmion is found. 
Over a narrow temperature interval, between 2.8 $J$ and 3.2 $J$ (corresponding to 950 - 1050 K), there is a steep increase in the average number of skyrmions 
to a saturation value of 4 skyrmions within the simulation cell. At even higher temperature (5 J and 7 J) the average number of skyrmions in the inherent structures, i.e. after minimisation, is unchanged.
Similar results are obtained by using the dis-LL algorithm but then the computational effort is more than an order of magnitude larger.

% ---------------------------------------------------------------------  Fig. 5  -------------------------------------------------

% ----------------------------------------------------------------------------------------------------------------------------------

\section{\label{sec:conclusion} Discussion}

By using the OSO algorithm, the task of minimising the energy on a curved manifold is transformed to a minimisation in linear space
making it possible to use the highly efficient LBFGS algorithm and an inexact line search.
This OSO-LBFGS algorithm
is found to provide an efficient method for finding local energy minima on the energy surface of magnetic systems,
outperforming in particular methods based on LL dynamics.
The method is used to study inherent structures of a ferromagnetic layer mimicking the widely studied Co/Pt(111) system as a function of temperature.
Over a remarkably narrow temperature range, the number of skyrmions in the inherent structures is found to rise abruptly to a high temperature saturation value.
These skyrmion configurations represent local minima on the energy surface that are favored over the global ferromagnetic energy minimum by entropy. 
In this system, the skyrmion configurations are separated by a low energy barrier from the ferromagnetic state, as has been reported previously~\cite{Lobanov2016}
so they are only expected to remain in the system after a fast quench to low temperature. This entropic preference for configurations involving skyrmions in the 
inherent structures is, however, likely a general property for spin systems and is consistent with the laser heating experiments where skyrmions were induced
in a quite different magnetic system~\cite{Vallobra2018}. 

The number of skyrmions can, in principle, be identified by evaluating the total topological charge of the finite temperature spin configuration and dividing by the topological charge of a single skyrmion~\cite{Ambrose2013}. The topological charge can be estimated as~\cite{Lohani2019}:
\begin{equation}
Q = \frac{\sum_{\Delta} \arctan \frac{\be_i \cdot (\be_j \times \be_k) }{ 1 + \left( \be_i \cdot \be_j + \be_i \cdot \be_k + \be_j \cdot \be_k  \right)} } {2\pi},
\end{equation}
where the sum is evaluated over triplets of neighboring spins. Such an approach can be useful at relatively low temperature where thermal fluctuations present only a small distortion of the minimum energy configuration. But, at high-temperature the magnetic configuration is highly disordered and can have an arbitrary value of the topological charge. Indeed, we have verified that the high temperature spin configurations generated here give essentially random values for the topological charge and are not related to the topological charge of the inherent structures at zero temperature. 

Nearly 200 minimization calculations have been carried out in the tests and the Co/Pt(111) simulations using the OSO-LBFGS method and in all cases convergence to a local energy minimum has been obtained without problems. The number of iterations needed increases as one would expect the more degrees of freedom are involved and the higher the initial temperature is. The calculations presented here indicate that the method is robust, but additional calculations in the future with application to other types of systems will further test and hopefully demonstrate its robustness and wide applicability.

The method has furthermore been applied to a Hamiltonian that describes more accurately itinerant electrons. The interaction in such systems often includes non-quadratic spin terms and cannot, therefore, be described by a Heisenberg type model. At each step of the energy minimization one needs to solve self-consistent field equations. We have shown that the minimization method presented here drastically reduces the number of SCF steps and we expect the algorithm to be useful in other self-consistent field calculations, such as density functional theory calculations.

The performance of the OSO-LBFGS algorithm could likely be improved further by
developing a preconditioner for the orthogonal optimization.
The dis-LL and damp-LL methods could be accelerated especially towards the end of a minimization where the gradient (i.e. effective field) is small by using a larger time step in an adaptive time step algorithm. An adaptive time step propagation based on Barzilian-Borwein rule has been used in micromagnetic simulations and found to give an acceleration by a factor of 2
~\cite{Exl2014}, but even with such acceleration, the dis-LL and damp-LL algorithms require significantly larger number of energy/gradient evaluations 
than the OSO-LBFGS algorithm. 

Some implementations make use of higher order integration methods\cite{Romeo2008,Leliaert2017}, such as 5th order Runge-Kutta method with adaptive time step. The application of this algorithm to the test problems presented here shows that the adaptive time step reduces the number of iterations needed for convergence, but the 5th order algorithm requires 6 evaluations of the gradient per iteration and as a result the total number of energy/gradient evaluations needed for convergence is larger than for the SIB algorithm.

The OSO algorithm should also be useful in other spin configuration optimizations, such as minimum energy path calculations using the geodesic nudged elastic band method,
which has previously been implemented with a VPO algorithm mimicking particle dynamics~\cite{Bessarab2015, Ivanov2020gneboso}.
Also, the algorithm could be used in searches for first order saddle points on energy surfaces for spin systems starting only from the initial state 
and without knowledge of any final state~\cite{Muller2018}.
From the saddle points, the rate of magnetic transitions can be estimated using harmonic transition state theory~\cite{Bessarab2012}
as has, for example, been done to estimate the lifetime of skyrmions as a function of temperature.~\cite{Uzdin2018,Bessarab2018}

The focus in this article is on the use of the OSO approach in optimisation,
but it could also be used in calculations of the dynamics of a spin system.
If one defines the $\vec{t}_{i}$ vector as 
$ \alpha \be_{i} \times \frac{\partial E}{\partial \be_{i}}  - \frac{\partial E}{\partial \be_{i}}  + \vec{f}_i $, 
where $\vec{f}_i$ is a random field, then the rotations can correspond to a trajectory obtained from the stochastic LL equation. 
In this way an algorithm for the integration of the stochastic LLG equation can be obtained. 
The use of exponential transformation for integration of the stochastic LLG equation has indeed been discussed by Lewis and Nigam.~\cite{Lewis2003}
An exact calculation of the rotation matrix has been used by Depondt and Mertens,~\cite{Depondt2009}
while an  approximation of the exponential transformation using Cayley transform~\cite{Lewis2003} is used in the method described by Mentik {\it et al}.~\cite{Mentink2010}

% ------------------------------------------------------------------------------------------------------------------------------------------------------------------------

\section{Acknowledgments}
We thank Julien Tranchida for helpful discussion and help with implementation of the algorithm in the LAMMPS software. 
Also, we thank Gideon M\"uller, Moritz Sallerman and Pavel Bessarab for helpful discussion and valuable comments on the manuscript. AVI thanks Filipp Rybakov for helpful discussions.
This work was funded by the Icelandic Research Fund, the University of Iceland Research Fund, 
and
the Russian Science Foundation (Grant 19-42-06302).
AVI is supported by a doctoral fellowship from the University of Iceland.

% ----------------------------------------------------------------------------------------------------------------------------------
%            Appendices
%
\appendix

\section{\label{app:OSO-SmallRot} Orthogonal Spin Optimization A}
Here the iteration process from Eq.~\eqref{eq: small_rotation} is described:
\begin{enumerate}
\item Set $k = 0$ and choose initial spin orientations $\{\be\,^{(k)}_i\}_{i=1..N}$. Let $\vec t_{i}\,^{(k)}$ be the torque acting on $i$th spin at $k$th iteration:
\begin{equation}
\vec t_{i}\,^{(k)} =  \be\,^{(k)}_{i} \times \frac{\partial E}{\partial \be\,^{(k)}_{i}}.
\end{equation}
Calculate the gradient vector:
\begin{equation}
 \vec{g}\,^{(k)} = \left(t_{0z}^{(k)}, -t_{0y}^{(k)}, t_{0x}^{(k)},
                          \dots, 
                          t_{Nz}^{(k)}, -t_{Ny}^{(k)}, t_{Nx}^{(k)}\right)^T
\end{equation}

\item Calculate initial search direction $\vec{p}\,^{(k)}$ according to the particular minimization algorithm chosen, 
see~\ref{app:qm}-\ref{app:lbfgs} for example.
\item Calculate the maximum magnitude of the torques: 
\begin{equation}
\Delta^{(k)} = \max_i  \left|\vec t_{i}\,^{(k)}\right|
\end{equation}
and set tolerance $\epsilon$ (for example, $\epsilon = 10^{-5}$ meV).
\item While $\Delta\,^{(k)}  > \epsilon$:
\begin{enumerate}
\item Compute $\lambda\,^{(k)}$ using a line search (for VPO algorithm $\lambda\,^{(k)}=1$) and calculate using~\eqref{eq:rotee}:
\begin{equation}
\be_i^{(k+1)} = e^{-\lambda^{(k)}P^{(k)}_i} \be^{(k)}_i, \forall i \in {1, 2, .., N}
\end{equation}
 where
skew-symmetric matrices 
\begin{equation}
P_{i}^{(k)} = 
\begin{pmatrix}
0 & p_{ix}^{(k)}& p_{iy}^{(k)}\\
-p_{ix}^{(k)} & 0& p_{iz}^{(k)}\\
 -p_{iy}^{(k)} & -p_{iz}^{(k)}& 0
\end{pmatrix} 
\end{equation} 
constructed using the search direction vector:
\begin{equation}
\vec{p}\,^{(k)} = \left(p_{0x}^{(k)}, p_{0y}^{(k)}, p_{0z}^{(k)},
                          \dots, 
                          p_{Nx}^{(k)}, p_{Ny}^{(k)}, p_{Nz}^{(k)}\right)^T
\end{equation}

\item Calculate new gradient $\vec{g}\,^{(k+1)} $ and $\Delta\,^{(k+1)} $. Calculate new search direction, $\vec{p}\,^{(k+1)} $, according to 
the particular minimization algorithm chosen, see Appendix~\ref{app:qm}-\ref{app:lbfgs} .  Set $k \leftarrow k + 1$   
\end{enumerate}
\item End
%Algorithm finished.
\end{enumerate}

\section{\label{app:OSO-LargeRot}  Orthogonal Spin Optimization B}
Here the iteration process from Eq.~\eqref{eq: large_rotation} is described.
The task is to minimize the
energy as a function of 
spin orientations, $E(\be_1,\be_2, .. \be_N)$. 
The $\be_i$ are parametrised using skew-symmetric matrices $\be_i = \mathrm{exp}(-A_i)  \be\,'_i$
\begin{equation}
A_i = 
\begin{pmatrix}
0 &a_{12}^{i}& a_{13}^{i}\\
-a_{12}^{i} &0 &a_{23 }^{i}\\
-a_{13}^{i} &-a_{23}^{i} &0  .
\end{pmatrix}
\end{equation}
It is enough to 
consider
only the upper diagonal part of $A_i$, and 
the energy is therefore a function of a $3N$-dimensional vector
\[\vec a = (a^{1} _{12}, a^{1} _{13}, a^{1} _{23},  ..., a^{i}_{12}, a^{i}_{13}, a^{i} _{23}, ..., a^{N} _{12}, a^{N} _{13}, a^{N} _{23}),\]
where the upper index, i, refers to the spin site. 
Let $F$ be
\begin{equation}
F(\vec a) =  E(e^{-A_1} \be_1', e^{-A_2} \be_2' \dots , e^{-A_N}\be_N')
\end{equation}

\vskip 0.3 true cm

% --------

{\bf Algorithm:}
\begin{enumerate}
\item Choose reference spin orientations $\{\be'_i\}_{i=1..N}$, 
set initial skew-symmetric matrices to zero, $\vec{a}^{(0)} = 0$, $\be_i^{(0)} = \be'_i $ and 
calculate initial gradient 
$\vec{g}\,^{(0)} = {\partial F}/{\partial \vec{a}\,^{(0)} }$.
\item Set $k = 0$ and calculate initial search direction $\vec{p}\,^{(0)}$ according to the particular minimization algorithm chosen, 
see~\ref{app:qm}-\ref{app:lbfgs}.
\item Set integer $u$ that counts the number of steps before updating the reference spins. For example, $u = 50$. 
\item Calculate the maximum magnitude of the torques: 
\begin{equation}
\Delta^{(k)} = \max_i  \left|\vec t_{i}\,^{(k)}\right|
\end{equation}
and set tolerance $\epsilon$ (for example, $\epsilon = 10^{-5}$ meV).
\item While $\Delta\,^{(k)}  > \epsilon$:
\begin{enumerate}
\item If $k\, \mathrm{mod} \, u = 0$ then update reference spins:
\begin{eqnarray}
&\be\,'_i  \leftarrow e^{-A\,^{(k)}_i} \be\,'_i, \forall i \in {1, 2, .., N}.\\
&\vec{a}\,^{(k)} = 0, \quad \vec{g}\,^{(k)} = {\partial E}/{\partial \vec{a}\,^{(k)} }, \\ 
&\text{calculate new } \vec{p}\,^{(k)}. 
\end{eqnarray}
\item Compute $\lambda\,^{(k)}$ using a line search (for VPO algorithm $\lambda\,^{(k)}=1$) and calculate:
\begin{eqnarray}
&\vec{a}\,^{(k+1)} = \vec{a}\,^{(k)}  + \lambda\,^{(k)}  \vec{p}\,^{(k)}, \\
&\be_i\,^{(k+1)} = e^{-A\,^{(k+1)}_i} \be\,'_i, \forall i \in {1, 2, .., N}
\end{eqnarray}
\item Calculate new gradient $\vec{g}\,^{(k+1)} $ and $\Delta\,^{(k+1)} $. Calculate new search direction, $\vec{p}\,^{(k+1)} $, according to 
the particular minimization algorithm chosen, see~\ref{app:qm}-\ref{app:lbfgs} .  Set $k \leftarrow k + 1$   
\end{enumerate}
\item End
%Algorithm finished.
\end{enumerate}
%
% -----------

\vskip 1 true cm
\section{\label{app:LineSearch} Choice of the step length.}
The step length parameter $\lambda$ is chosen in such a way
that the strong Wolfe conditions~\cite{Nocedal2006} and/or approximate Wolfe conditions~\cite{Hager2006} are satisfied
%
% \begin{widetext}
\begin{eqnarray}
\label{eq: strw1}
& F(\vec {a} + \lambda\, \vec {p}\,) \le F(\vec {a}) + c_1 \lambda\, \nabla_{\vec {a}} F(\vec {a}) \cdot \vec {p}\\
\label{eq: strw2}
& |\nabla F(\vec {a} + \lambda \vec{p}) \cdot \vec{p}\,|  \le c_2 | \nabla_{\vec{a}} F(\vec {a}) \cdot \vec {p}\,|
\end{eqnarray}
and/or
\begin{eqnarray}\label{eq: aprw}
&F(\vec {a} + \lambda \vec {p}\,)  \le F(\vec {a}\,) + \epsilon |F(\vec {a}\,)| \\
&(2\delta - 1)\nabla_{\vec{a}} F(\vec {a}\,) \cdot \vec {p}\, \ge 
\nabla_{\vec{a}} F(\vec {a} + \lambda \vec {p}\,) \cdot \vec {p} \ge
 \sigma \nabla_{\vec{a}} F(\vec {a}\,) \cdot \vec {p}
\end{eqnarray}
% \end{widetext}
%
with $0<c_1<c_2<1, \epsilon>0, \delta<\mathrm{min}{.5, \sigma}, \sigma<1$.
The parameters were chosen according to Ref.~\cite{Nocedal2006} and \cite{Hager2006} as
\begin{equation}
c_1 = 10^{-4}, c_2 = 0.9, \delta = 0.1, \sigma =0.9, \epsilon=10^{-6}
\end{equation}
It can be shown that after several iterations, a step length of 1 guarantees satisfaction of the strong Wolfe conditions in the LBFGS algorithm,~\cite{Nocedal2006} and therefore, a trial step of $\lambda=1$ is always used first to test these conditions. If they are not satisfied, an inexact line search procedure based on the cubic interpolation is used.~\cite{Nocedal2006} Approximate Wolfe conditions~\cite{Hager2006} are always examined at the minimum of the cubic interpolation (see~\ref{app:cubic_f}).

The other approach for the choice of the step length is based on the maximum rotation of the spins as described in the main text.
% ----------------------------------------------------------

% Appendix D

\section{\label{app:qm} VPO algorithm}

Values of two parameters need to be chosen, 
the time step $\Delta t$ and 
an effective (artificial) mass, $m$.
Since the system accelerates when the gradient points in a similar direction in subsequent iterations,
the time step can be chosen to be small,
%. The  
$\Delta t = 0.005$. 
The effective mass was chosen to be $m = 0.01$.
At the $k$-th iteration the search direction is chosen according to the following:\\

if k = 0: 
\begin{align}
&\vec{v}\,^{(k)} = 0\\
&\vec{p}\,^{(k)} = -\frac{\vec{g}\,^{(k)} \Delta t^2}{2m} \\
&\mathrm{return } \, \vec{p}\,^{(k)} .
\end{align}

else:
\begin{align}
&\vec{v}\,^{(k)} = \vec{v}\,^{(k-1)} - \frac{1}{2} \left(  \vec{g}\,^{(k-1)} + \vec{g}\,^{(k)}\right) \Delta t / m\\
&\beta^{(k)} = \vec{g}\,^{(k)} \cdot \vec{v}\,^{(k)} / { \vec{g}\,^{(k)} \cdot \vec{g}\,^{(k)}} ,\\
&\mathrm{if\,\,} \beta\,^{(k)} > 0 \mathrm{\,\,then\,\,set\,\,} \beta\,^{(k)} = 0.\\
&\vec{v}\,^{(k)} \leftarrow \beta^{(k)} \vec{g}\,^{(k)},\\
&\vec{p}\,^{(k)} = \vec{v}\,^{(k)} \, \Delta t -\frac{\vec{g}\,^{(k)} \Delta t^2}{2m},\\
&\mathrm{return } \, \vec{p}\,^{(k)} .
\end{align}
%----------------------------------------------------------------------------------------------------------------------------------------------

\section{\label{app:CG} CG algorithm}
The Fletcher–Reeves nonlinear conjugate gradient method~\cite{Nocedal2006} is given here for completeness of the paper.
At the $k$-th iteration the search direction is chosen according to the following

if k = 0: 
\begin{align}
&\vec{p}\,^{(k)} = -\vec{g}\,^{(k)} \\
&\mathrm{return } \, \vec{p}\,^{(k)} .
\end{align}

else:
\begin{align}
&\beta^{(k)} = \frac{\left| \vec{g}\,^{(k)} \right|^2 }{ \left| \vec{g}\,^{k-1} \right|^2},\\
&\vec{p}\,^{(k)} = \beta^{(k)} \vec{p}\,^{(k)} - \vec{g}\,^{(k)},\\
&\mathrm{return } \, \vec{p}\,^{(k)} .
\end{align}

% ------------------------------------------------------------------------------------------------------------------------------------------------
\section{\label{app:lbfgs} L-BFGS algorithm}
Limited-memory Broyden-Fletcher-Goldfarb-Shanno (L-BFGS) algorithm~\cite{Nocedal2006} is given here for completeness.

\vskip 0.3 true cm

% --------

{\bf Algorithm:}
Set $k = 0$. Let $m$ be amount of previous step which are stored in memory.
\begin{enumerate}
\item[(1)] Calculate $n = k\, \mathrm{mod} \, m$
\item[(2)]  {\bf if} k = 0: 
\begin{eqnarray}
& \vec{p}\,^{(k)} = \vec{g}\,^{(k)},\\
& \vec{d}_{j} = 0,\, \vec{y}_{j} = 0,\,\rho_j=0\, \forall j \in {0, 1 .. m-1} \\
% & \vec{d}_{n} = 0,\\
% & \vec{y}_{n} = 0,\\
&\textbf{end if} \nonumber
\end{eqnarray}
\textbf{else}:
\begin{eqnarray}
& \vec{d}_{n} = \lambda^{(k-1)}\vec{p}\,^{(k-1)},\\
& \vec{y}_{n} = \vec{g}\,^{(k)} - \vec{g}\,^{(k-1)}, \\
% & sy = \vec{s}_{n}\,^{(k)} \cdot \vec{y}_n\,^{(k)},\\
% & \rho_n = 1/ys,\\
& \rho_n = 1/ \left(\vec{d}_n \cdot \vec{y}_n\right),\\
& \textbf{if } \rho_n < 0 \text{ set } k=0 \text{ and go to (1) }  \textbf{end if}\\
& \vec{q} = \vec{g}\,^{(k)}
\end{eqnarray}
\qquad \textbf{for} $l$ in $(m - 1), (m - 2) \dots 0$:
\begin{eqnarray}
& \qquad j = (l + n + 1) \,\mathrm{mod}\ m,\\
& \qquad \gamma_j = \rho_j \left(\vec{d}_{j}\cdot \vec{q}\,\right),\\
& \qquad \vec{q} \leftarrow \vec{q} - \gamma_j \vec{y}_{j}
% ,\\
% & \textbf{end for} \nonumber
\end{eqnarray}
\qquad \textbf{end for}
\begin{eqnarray}
& \qquad \vec{p}\,^{(k)} = \vec{q} / (\rho_n \left(\vec{y}_{n} \cdot \vec{y}_n\right) ) 
\end{eqnarray}
\qquad \textbf{for} l in $0, 1, \dots (m - 1)$:
\begin{eqnarray}
& \qquad \textbf{if}\, k < m \text{ then } j = l,  \textbf{end if}\\
& \textbf{else}\, j = (l + n + 1) \,\mathrm{mod}\ m, \textbf{end else}\\
& \qquad \vec{p}\,^{(k)} \leftarrow \vec{p}\,^{(k)} + \vec{d}_{j} \left( \gamma_j - \rho_j \left(\vec{y}_{j} \cdot \vec{p}\,^{(k)}\right)\right)
% \\
% & \textbf{end for} \nonumber
\end{eqnarray}
 \qquad \textbf{end for} \\
\textbf{end else}
\item[(3)] \textbf{return -$\vec{p}\,^{(k)}$}
\end{enumerate}

%------------------------------------------------------------------------------------------------------------------------------------------------

\section{\label{app:cubic_f} Cubic interpolation}

Let $f$ be defined on $[0, r]$ and $f(0), f(r), f'(0), f'(r)$ be known. Then the cubic interpolation of $f$ is:
\begin{eqnarray}
&f(\alpha)^{{\rm approx}} = c_1\alpha^3 + c_2\alpha^2 + c_3 \alpha + c_4 \\
&c_1 =  -  \frac{2f(r) - 2f(0)} {r^3} +  \frac{f'(r) + f'(0)}{r^2},\\
&c_2 =  \frac{3f(r) - 3f(0)}{r^2} - \frac{f'(r) + 2 f'(0)}{r},\\
&c_3 =  f'(0), \quad c_4 =  f(0) 
\end{eqnarray}
and the minimum of this cubic function is at
\begin{equation}
\alpha_0 =  (-c_2 + \sqrt{c_2^2 -3c_1c_3}) / 3c_1
\end{equation}
or at the boundaries of the interval $[0, r]$.

% ------------------------------------------------------------------------------------------------------------------------------------------------

\section{\label{app:um} Orthogonal matrix and gradient of the energy}

Let $A$ be
\begin{equation}
A = 
\begin{pmatrix}
0 & a & b \\
-a & 0 & c \\
-b & -c & 0 \\
\end{pmatrix}
\end{equation}
Let $\theta = \sqrt{a^2 + b^2 + c^2}$. Then the eigenvalues are:
\begin{eqnarray}
&\lambda_1 = 0, \\
&\lambda_2 = -i \theta,\\
&\lambda_3 = i \theta.
\end{eqnarray}
The eigenvectors are:
\begin{equation}
\hat{v}_1 = \frac{1}{\theta}
\begin{pmatrix}
c \\
-b\\
 a
\end{pmatrix},\quad
\hat{v}_2 = \frac{1}{\theta \sqrt{2(a^2 + c^2)}}
\begin{pmatrix}
bc + ia\theta\\
a^2 + c^2\\
ab - ic\theta
\end{pmatrix}, \quad
\hat{v}_3 = \hat{v}_2\,^*
\end{equation}
if $a = 0$ and $c = 0$ then 
\begin{equation}
\hat{v}_2 = \frac{1}{\sqrt{2}}
\begin{pmatrix}
\mathrm{sign}(b) i \\
0.0\\
1.0
\end{pmatrix}
\end{equation} 
The matrix exponential can 
then
be calculated 
%then 
as:
\begin{equation}
e^{A} = V L V^{\dagger},
\end{equation}
where $L$ is a diagonal matrix, $\mathrm{diag}(L) = (1, e^{-i\theta}, e^{i\theta}) $, and $V$ is a unitary matrix, columns of which are eigenvectors of $A$.
The gradient of the energy is then $g_{\alpha\beta} = {\partial E}/{\partial a_{\alpha\beta}}$:
\begin{equation}
G = V\left(\left(V^{\dagger} T V\right) \circ D\right)V^{\dagger},
\end{equation}
where 
$\circ$ denotes Hadamard product,
\begin{equation}
D_{\alpha\beta} = \frac{ e^{\lambda_\alpha - \lambda_\beta}- 1}{\lambda_\alpha - \lambda_\beta},
\end{equation}
and
\begin{equation}
T =
\begin{pmatrix}
0& t_{z}& -t_{y}\\
-t_{z}& 0& t_{x}\\
t_{y}& -t_{x}& 0
\end{pmatrix}, \quad \vec t =  \hat{s} \times \frac{\partial E}{\partial \hat{s}}.
\end{equation}
\\
 
\textit{Rodrigues’ formula for rotations.}\\
Let $q = \cos\theta, w =  1 - \cos\theta$ and $x = a/\theta, y=b/\theta, z=c/\theta$,  $s_1= -y z w$,  $s_2 = x z w$, $s_3=-x y w$, $p_1 = x \sin\theta$,  $p_2 = y \sin\theta$,  $p_3 = z \sin\theta$ then the matrix exponential can be calculated as:
\begin{equation}\label{eq:rf}
\exp(A) = 
\begin{pmatrix}
q + z^2 w & s_1 + p_1& s_2 + p_2\\ 
s_1 - p_1&q + y^2 w& s_3 + p_3\\
s_2 - p_2& s_3 - p_3 &q + x^2 w
\end{pmatrix}
\end{equation}

%
% The \nocite command causes all entries in a bibliography to be printed out
% whether or not they are actually referenced in the text. This is appropriate
% for the sample file to show the different styles of references, but authors
% most likely will not want to use it.
%\nocite{*}

% ------------------------------------------------------------------------------------------------------------------------------------------------
\newpage
%\bibliography{references}% Produces the bibliography via BibTeX.

\begin{thebibliography}{00}
\bibitem{random} The azimuth angles and cosines of polar angles of spin configurations were generated using numpy.random.uniform(0, 2$\pi$) and numpy.random.uniform(-1, 1) functions (Numpy 1.15.2), respectively, with the following seeds:
187, 451, 559, 661, 1003, 1033, 1195, 1321, 2533, 3091, 
3169, 3211, 3955, 4033, 4585, 5029, 5143, 5287, 5779, 6109,
6373, 6727, 7327, 7681, 7873, 7915, 7939, 7963, 8485, 8713, 
8845, 8899, 9079, 9121, 9295, 9337, 9493, 9817, 10159, 10309.
\end{thebibliography}

\end{document}